\documentclass{article}

\usepackage[english]{babel}
\usepackage[letterpaper,top=2cm,bottom=2cm,left=3cm,right=3cm,marginparwidth=1.75cm]{geometry}
\usepackage{graphicx}
\usepackage{amsmath}
\usepackage[version=4]{mhchem}
\usepackage{gensymb}
\usepackage{siunitx}
\usepackage{tabularx}
\usepackage{setspace}
\usepackage{makecell}
\usepackage{ragged2e}
\usepackage{titlesec}
\usepackage{float}
\usepackage[utf8]{inputenc}
\usepackage[T1]{fontenc}
\usepackage{newunicodechar}
\newunicodechar{₂}{$_2$}
\usepackage{subcaption}
\titleclass{\subsubsubsection}{straight}[\subsection]
\usepackage{csquotes}
\usepackage[backend=biber,style=nature,autopunct=true,maxnames=5]{biblatex}
\usepackage{xurl}

\usepackage[colorlinks=true,allcolors=blue]{hyperref}

\newcounter{subsubsubsection}[subsubsection]
\renewcommand\thesubsubsubsection{\thesubsubsection.\arabic{subsubsubsection}}

\titleformat{\subsubsubsection}{\normalfont\normalsize\bfseries}{\thesubsubsubsection}{1em}{}
\titlespacing*{\subsubsubsection}{0pt}{3.25ex plus 1ex minus .2ex}{1.5ex plus .2ex}
\makeatletter
\renewcommand\paragraph{\@startsection{paragraph}{5}{\z@}{3.25ex \@plus1ex \@minus.2ex}{-1em}{\normalfont\normalsize\bfseries}}
\renewcommand\subparagraph{\@startsection{subparagraph}{6}{\parindent}{3.25ex \@plus1ex \@minus .2ex}{-1em}{\normalfont\normalsize\bfseries}}
\def\toclevel@subsubsubsection{4}
\def\toclevel@paragraph{5}
\def\l@subsubsubsection{\@dottedtocline{4}{7em}{4em}}
\def\l@paragraph{\@dottedtocline{5}{10em}{5em}}
\def\l@subparagraph{\@dottedtocline{6}{14em}{6em}}
\makeatother
\DeclareSourcemap{
  \maps[datatype=bibtex]{
    \map{
      \step[fieldsource=doi,final]
      \step[fieldset=url,null]
      \step[fieldset=urldate,null]
    }
  }
}

\begin{document}

\begin{center}
    {\large\bfseries Monolithic Barium Titanate Nanophotonics and Electro-optics}
    
    \vspace{1em}
    
    Sarah Berman,$^{1}$ Sina Dereshgi,$^{1}$ and David Barton$^{1}$
    
    \vspace{0.8em}
    
    {\small $^{1}$ Department of Materials Science and Engineering,\\
    Northwestern University, Evanston, IL 60208, USA}
    
    \vspace{0.5em}
    
    {\small Corresponding author: sarahberman2028@u.northwestern.edu}
    
    \vspace{2em}
    
    \textbf{Abstract}
    
    \vspace{0.5em}
    
    \begin{minipage}{0.85\textwidth}
    \small
%%====================================================================
Barium titanate-on-insulator (BTOI) is a compelling material for high-speed integrated photonic modulators due to its large Pockels coefficient ($r_{42}$ > 1200 pm/V in bulk), which allows for the miniaturization of modulators while maintaining strong electro-optic performance. Sub-wavelength nanostructures monolithically etched into BTOI are particularly exciting, as they offer a path toward subwavelength light-matter interaction and reduced modulator energy consumption. Here, we design, fabricate, and characterize monolithic one-dimensional nanophotonic crystals (PhCs) and high-Q (230k) photonic-crystal Fabry–Pérot cavities in BTOI. We develop and optimize a nanofabrication process that yields anisotropic (75$^\circ$ sidewalls) and deep etching that features low optical loss, with racetrack resonators achieving intrinsic quality factors near 1 million and propagation losses of about 0.5 dB/cm. Our photonic crystals exhibit bandgap contrasts greater than 40 dB, and Fabry–Pérot cavities reach loaded quality factors up to 230k. We verify ferroelectric domain alignment via second-harmonic generation microscopy and extract an effective Pockels coefficient of \~154 pm/V from DC electro-optic tuning. By probing the microwave response at the PhC band edge, where modulation bandwidth is set by the material's electro-optic response rather than cavity photon lifetime, we measure a 3-dB electro-optic bandwidth of 11 GHz and a 6-dB bandwidth of 21 GHz, consistent with the frequency-dependent roll-off of BTO's $r_{42}$ coefficient near 10 GHz. Finally, we show a variety of modulation effects in resonators and at photonic crystal band edges, including sideband-resolved modulation, resonant bandwidth-limited modulation, and photonic-crystal based single sideband modulation and frequency comb generation. Beyond modulation, the low-loss, high-contrast BTOI nanostructures demonstrated here open the door for future visible-wavelength electro-optic devices, dispersion engineered cavities and waveguides for nonlinear optics, and compact quantum photonic components. 
\end{minipage}
\end{center}
\vspace{1em}
\begin{spacing}{1}
%%====================================================================
\section{Motivation}
Electro-optic (EO) integrated photonic devices are important for low energy consumption and broadband modulation of optical signals, relevant for applications in optical communications (\cite{OpticalCommIntro}), microwave photonics \cite{MicrowaveIntro}, advanced computation \cite{AdvancedComputeIntro}, and quantum information processing \cite{QuantumIntro}. Achieving high speed (>10 GHz), low drive voltage (<1V peak-to-peak), and compact footprint in a single modulator in CMOS-compatible formats is required for future scalable integrated photonic computing and communications systems \cite{ref52}. Of the approaches employed for refractive-index based modulation, the Pockels effect in non-centrosymmetric materials is particularly compelling because of its broadband index modulation without inducing additional loss \textit{via} phenomena like carrier injection\cite{PhotonicsForComputing}.  Photonic crystals (PhCs) are a popular building block for compact nanostructured devices and have been fabricated in a variety of platforms like silicon \cite{SiWaveguide}, GaAs \cite{GaAsWaveguide}, SiN \cite{SiNWaveguide}, and LiNbO$_3$  \cite{ref45, LNNanoStructures1, LNNanoStructures2}. PhCs allow for flexible and precise control of the flow of light on a chip \cite{WhySlowLight}\cite{SlowlightPhCs}, including subwavelength localization of light and dispersion engineering \cite{ref41}. This has enabled a myriad photonic devices and exploration of new phenomena, including polarization multiplexers \cite{PolarizationMultiplexer}, topological photonic edge states \cite{TopologicalEndStates}, femtosecond pulse dispersion engineering \cite{yu2022integrated,du2020silicon}, and strong coupling to quantum emitters \cite{QuantumEmitter,ohta2011strong,ding2025purcell}. Additionally, PhC-based structures are useful for electro-optic modulators, as low mode volume cavities \cite{ref45} and Fabry–Pérot cavities \cite{ref43, FPFreqComb} can significantly reduce the required bit switching energy by reducing the drive voltage and device capacitance. 

Barium titanate (BTO) is a compelling platform for electro-optic devices because of its high (r\textsubscript{42}>1200 pm/V in bulk \cite{ref56}) Pockels coefficient and CMOS compatibility. Recent demonstrations of high quality wafer-scale film growth and device fabrication show promise \cite{ref17,ref19, ref57} as a competing platform to lithium niobate on insulator. A majority of devices rely on barium titanate-on-insulator (BTOI) as an active layer under SiN or Si \cite{ref14,ref30} ridge waveguides, including resonators and interferometers \cite{ref17,ref19}. However, this limits the electro-optic effect in these devices, as the mode is primarily localized in a passive material with similar or higher refractive index. Monolithic integration of waveguides and nanophotonic structures therefore promises significant improvements in device performance by co-locating the electro-optic material and optical mode \cite{ref19,prountzou2026electro}. Together, this means that monolithic BTOI devices can potentially out-perform lithium niobate-on-insulator (LNOI) \cite{ref59} or lithium tantalate\cite{LithiumNiobatePhotonics} devices in terms of energy consumption and size\cite{ref52}.  Part of the success of LNOI as an integrated photonics platform is enabled not only by its EO coefficient, but by the ability to create high-fidelity, sub-wavelength nanostructures \cite{ref45, LNNanoStructures1, LNNanoStructures2} that enhance light-matter interactions to reduce the overall size of the device, and therefore improve modulation performance through the reduction of device capacitance \cite{CapacitanceCitation}. This has proved challenging in BTOI, as low-loss monolithic etching of nanostructures has not been well established. There are few demonstrations of wavelength-scale monolithic devices in this platform, including Mach-Zehnder interferometers (MZIs) and resonant microring or racetrack modulators, that feature low optical loss \cite{ref17}. Fabricating low loss and monolithic photonic crystals (PhC) in BTOI provides this missing link \cite{ref51} that can provide energy efficient and compact photonic devices in this platform.

Here, we demonstrate monolithic nanophotonic and electro-optic devices in thin film BTOI with low loss and high tuning efficiency.  We develop a dry-etching fabrication process that achieves anisotropic sidewalls (75$^\circ$) and low roughness sufficient for low-loss nanostructures. We evaluate fabrication quality using ring and racetrack resonators, achieving propagation losses of about 0.5 dB/cm and intrinsic Q factors over 1 million in the near infrared. We then design and fabricate 1-D PhCs in BTOI, demonstrating geometry-tunable bandgaps with contrasts exceeding 40 dB, and use PhC mirrors to construct Fabry--P\'{e}rot cavities with loaded Q factors up to 230k. After verifying ferroelectric domain alignment via confocal second harmonic generation microscopy, we measure an effective Pockels coefficient of \textasciitilde150 pm/V in optical resonators. We further characterize the microwave electro-optic response using the PhC band edge, measuring a 3-dB bandwidth of 11 GHz and a 6-dB bandwidth of 21 GHz. Finally, we demonstrate asymmetric sideband generation near a photonic crystal band edge, frequency comb generation under strong microwave driving, and investigate the microwave response of optical resonators in both the sideband resolved and unresolved regimes.  

\subsection{BTO Nanofabrication}
\begin{figure}[h]
    \centering

    \includegraphics[width=\textwidth]{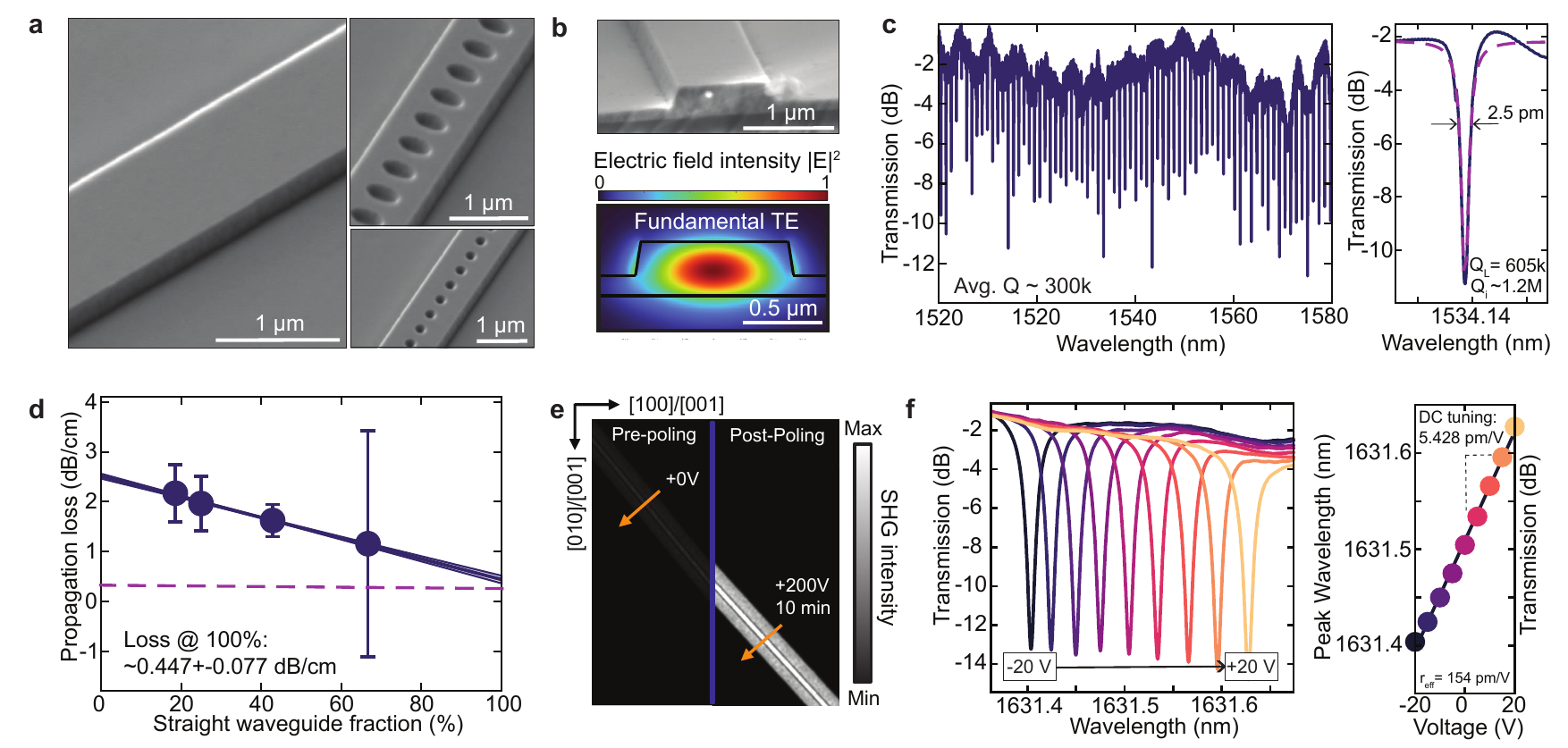}
    
    \caption{a) SEM images of fabricated structures demonstrating smooth, redeposition-free sidewalls with partially-etched elliptical sections in the photonic crystals. b) Cross-sectional SEM image showing approximately a 75-degree sidewall angle and TE-polarized MODE simulation in lumerical (99\% TE) c) Left: Transmission spectrum of a typical ring resonator with an average loaded quality factor of 300k. Right: Example high-Q resonance from the same resonator, with intrinsic Q of ~1.2M. The racetrack resonator has a straight length of 249 microns and bending radius of 100 microns. d) Propagation loss vs. \% straight length for racetrack resonators on the same chip with a bend radius of 100 um with error bars showing the spread of calculated loss values at each straight length. The straight waveguide loss is extracted by extrapolating the slope of the propagation loss to 100\% straight fraction. e) SHG microscopy image before and after poling of a 45$^\circ$- oriented waveguide showing maximal SHG intensity across portion of poled waveguide. f) Electro-optically tuned racetrack resonance to measure effective Pockels coefficient. Extracted resonance position as a function of voltage corresponds to a tuning efficiency of 5.428 pm/V, corresponding to an effective Pockels coefficient of 154 pm/V.
}
    \label{fig:FIG1}
\end{figure}

Realizing low-loss PhCs in BTO requires fabrication quality sufficient to sustain low per-period scattering. A high sidewall angle, low surface and sidewall roughness, and high-fidelity mask transfer are necessary to maintain effective refractive index contrast between periods while still confining a propagating mode. Achieving a high degree of anisotropy in our etch is essential for photonic crystal performance. For example, a 60° sidewall angle in a 200-nm partial etch would reduce the effective fill fraction of the air holes by approximately 30\%, substantially degrading the index contrast and bandgap depth of the PhC. We start with 330-nm thick BTO-on-insulator (La Luce Cristallina). We define single mode waveguides with widths of 1-1.4 $\mu$m top widths and photonic crystal holes with feature sizes of approximately 100 nm using electron beam lithography. Following reactive ion etching using Ar+ and $CHF_3$ gases, the mask and redeposition are stripped, and devices with active components are metallized using photolithography (see Methods section for fabrication details). Figure \ref{fig:FIG1}a shows SEM images of single mode waveguides and photonic crystal waveguides that feature low roughness and high quality etching of nanostructures with length scales of approximately 100 nm using this approach. A representative cross-sectional SEM image (Figure \ref{fig:FIG1}b) of the waveguide shows an etch angle of approximately 75$^\circ$. Using this, we designed single TE mode waveguides, where a majority of the optical mode is contained within the BTO, as shown in the bottom of Figure \ref{fig:FIG1}b.
We first investigate the quality of our fabrication process by measuring ring and racetrack resonators. Across rings with bus-waveguide widths of 1-1.2 \textmu m and separations between 0.4~\textmu m and 1.3~\textmu m, we identify a critically coupled gap of 0.85 \textmu m for a waveguide width of 1 \textmu m, at which the average Q-factor of a device no longer increases with increasing distance from the bus waveguide. (Figure \ref{fig:FIG1}c) shows a representative transmission spectrum of a device near critical coupling, where we extract a maximum loaded $Q_L$ of 605k and an intrinsic $Q_i$ of over 1 million (Figure \ref{fig:FIG1}c). 

We separately extract the propagation loss by varying the straight-section length of racetrack resonators with a fixed 100~\textmu m bend radius and extrapolating a linear fit extended to the $100\%$ straight case. The average propagation losses for these experiments are shown in Figure \ref{fig:FIG1}d, and the slope yields a straight waveguide propagation loss of 0.447 $\pm$ 0.077 dB/cm (Figure \ref{fig:FIG1}d) (see Methods). These values are on par with recent reports for monolithic BTO~\cite{ref17} and, critically, confirm that the sidewall quality and etch profile of our process are sufficient to support the high-aspect-ratio periodic features required for low-loss 1-D photonic crystals, where scattering at each etched period would otherwise compound into substantial transmission loss and degraded bandgap contrast.

For commercially available a-b oriented films, BTO's effective electro-optic coefficient is maximized when an applied electric field and optical mode polarization are along the $y$-$z$ plane \cite{ref56} of the crystal, which allows for access to the $r_{42}$ component of its EO tensor. Previous demonstrations have shown effective Pockels coefficients $r_{eff}\approx 130-160$ pm/V range \cite{ref57,ref17}. Following etching, we pole our devices and verify this using second-harmonic generation (SHG) microscopy, using the nonlinear generation intensity as a proxy for ferroelectric domains alignment. During \textit{in situ} poling, we observe the SHG intensity rise monotonically with applied bias and saturate once the domains are aligned along the field direction (Figure \ref{fig:FIG1}e). We find that applying 200 V across a 6 micron gap for 10 minutes will fully pole our samples. The poled state appears stable if no other stimuli are applied post poling. SHG images acquired 6 weeks after initial poling show only small changes in intensity. We additionally find that applying voltages to this device greater than $\pm20$ V will begin to reorient domains, evidenced by drift in the DC tuning response under randomized voltage sequences (See Supplementary Fig. 2). This window, set by the coercive field of the film at our contact spacing, defines the operating range within which the device behaves as a stable linear modulator.

We extract the effective Pockels coefficient for these devices by electro-optically tuning the resonant wavelength of a racetrack resonator. We perform tuning experiments across voltage ranges of $-20$~V to $20$~V (-3.3$\times10^{6}$ V/m to 3.3$\times10^{6}$ V/m) (Figure \ref{fig:FIG1}f). We find a linear dependence on the resonant frequency expected from the electro-optic effect. We extract the slope of our DC tuning curve (i.e. the resonant wavelength shift) and normalize it by the FSR of our resonator to extract the effective Pockels coefficient using \cite{ref58}:
\begin{equation}
r_{\text{eff}} = \frac{\lambda_0 \cdot L_{\text{gap}} \cdot 2 \cdot \frac{d\lambda}{dV}}{n_{\text{eff}}^3 \cdot \Gamma_{\text{eo}} \cdot L_{\text{probe}} \cdot \text{FSR}}
\label{eq:reff_extraction}
\end{equation}
where $\lambda_0 = 1550$~nm, $L_{\text{gap}} = 6$~\textmu m, $d\lambda/dV = 5.428$~pm/V (DC tuning slope), $n_{\text{eff}} = 1.9$, $\Gamma_{\text{eo}} = 0.51$ (see Supplementary Fig. 3), $L_{\text{probe}} = 134$~\textmu m, and FSR~$= 1.1$~nm (see SI). This yields an effective Pockels coefficient of 154~pm/V. This is comparable to existing literature values \cite{ref19}\cite{ref56}\cite{ref39} and is significantly larger than the maximal accessible Pockels coefficient in comparable material platforms such as lithium niobate~\cite{ref59} and lithium tantalate~\cite{ref59}. 

\subsection{BTO PhC Device Design and Characterization}
\begin{figure}[h]
    \centering
    
       \includegraphics[width=\textwidth]{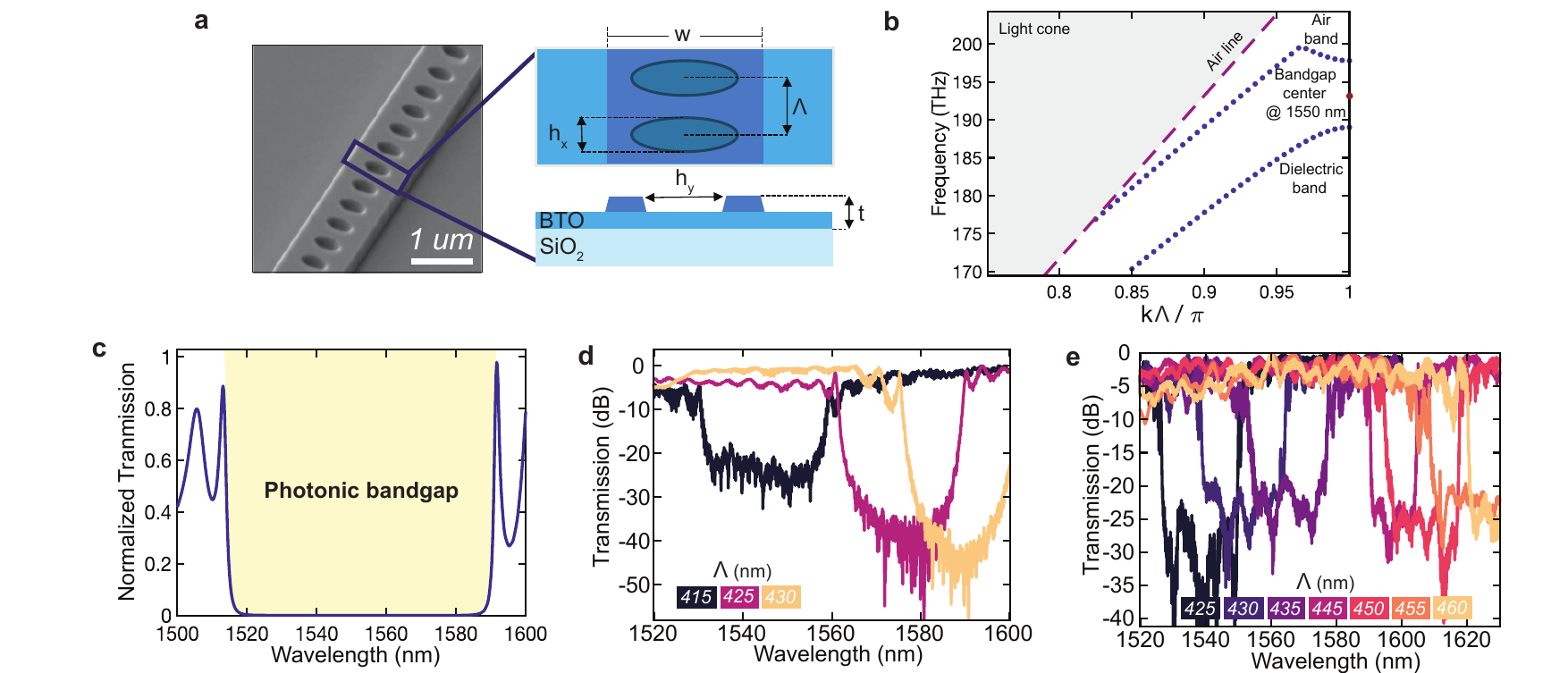}
        \caption{a) SEM image and schematic of a BTO photonic crystal with deliniated design parameters. The BTO PhC waveguide has a width of \textit{w}=1 um, layer height of \textit{t}=330nm, lattice constant (period) of a ranging from 420 nm to 450 nm and a partially etched layer with a slab thickness of 130 nm. The elliptical hole has dimensions of \textit{h}\textsubscript{\textit{x}} between 260 nm and 400 nm, and \textit{h}\textsubscript{\textit{y}} between 260 nm and 600 nm, and a partially etched depth of 200 nm. b) Dispersion properties of a designed PhC simulated in COMSOL through eigenfrequency analysis. The dark purple dots represent the air and dielectric bands, with a bandgap centered around 1550 nm. The light light is shown in dotted purple to represent where propagation is disallowed. c) FDTD simulated bandgap centered at 1550 nm for a photonic crystal with \textit{h}\textsubscript{\textit{x}}=300 and \textit{h}\textsubscript{\textit{y}}=600 with $\Lambda$ = 0.445 um, showing a bandgap width of 70 nm. d) Characterized PhCs with shifting period demonstrating high contrast, geometrically shiftable bandgaps. Unit cells have $h_x$=$h_y$=260 nm with a height of 407 nm on a slab of 200 nm BTO. e) Characterized PhCs demonstrating a wider period sweep; each PhC has 400 unit cells with elliptical hole dimensions \textit{h}\textsubscript{\textit{y}}= 500 nm,  \textit{h}\textsubscript{\textit{x}}=200 nm with a 210 nm height  on a 110 nm slab of BTO.}.
    \label{fig:FIG2}
\end{figure}
Our anisotropic sidewalls allow for an effective refractive index contrast between etched unit cells sufficient to open a wide bandgap while maintaining low scattering loss. Figure \ref{fig:FIG2}a shows an SEM of a typical photonic crystal, with a schematic of the photonic crystal design on the right. We design our PhCs on a BTO wafer comprised of a 330~nm thick BTO layer, a 3 \textmu m thick silica layer, and a 500 \textmu m thick silicon substrate. We design photonic crystals that are partially etched through the BTO layer to ensure appropriate optical mode confinement while maintaining the ability for high electric field penetration through the photonic crystal. We parametrically sweep unit cell dimensions of etched air ellipses with a major axis between 0.13~\textmu m and 0.3~\textmu m and a minor axis between 0.13~\textmu m and 0.18~\textmu m, with a slab thickness of 200~nm and unit cell height of 400~nm. Figure \ref{fig:FIG2}b shows the band structure for a photonic crystal waveguide with unit cell dimensions of \textit{h}\textsubscript{\textit{x}}=300 and \textit{h}\textsubscript{\textit{y}}=600 with $\Lambda$ = 0.445 um on a BTO slab thickness of 200 nm with a partial etch depth of 200 nm, which features a wide optical band gap of approximately 70 nm due to the elliptical holes that maximize index contrast in this configuration. Figure \ref{fig:FIG2}c shows a three-dimensional finite-difference time domain (FDTD) simulation (ANSYS Lumerical) of the transmission spectrum through this photonic crystal, showing a clear optical band gap in the C-band.

We experimentally validate this by fabricating photonic crystals with differing periods to show the geometrically tunable photonic band gap. We fabricate photonic crystals on both 330 nm thick and 400 nm thick BTO thin films, the two thicknesses of commercial BTO wafer we were able to obtain. Figure \ref{fig:FIG2}d) shows photonic crystals optimized for high contrast, showing greater than 40 dB contrast using a sufficiently high number of unit cells (500) and unit cell dimensions of h\textsubscript{\textit{x}}=h\textsubscript{\textit{y}}=260 and a unit cell height of 407 nm with a BTO slab height of 200 nm. We additionally implement a taper with regards to major and minor axis length of our elliptical holes to reduce scattering at the waveguide-photonic crystal interface~\cite{ref43,witmer2020silicon}. Figure \ref{fig:FIG2}e shows devices with periods ranging from 425 to 460 nm with unit cell dimensions of h\textsubscript{\textit{x}}=200 nm and h\textsubscript{\textit{y}}=500 nm, with partial etch depths of 200 nm on a slab of 130 nm BTO, which yield band gaps that span the entire C and L telecommunications bands. Together with the tunability of band gap center and width, this gives us a geometrically configurable design space from which we select the mirrors for the Fabry--P\'{e}rot cavities and sharp band edges used for microwave modulation~\cite{ref61}. 

\subsection{BTO Fabry--P\'{e}rot Characterization and Tuning}
\begin{figure}[htp]
    \centering
        \includegraphics[width=\textwidth]{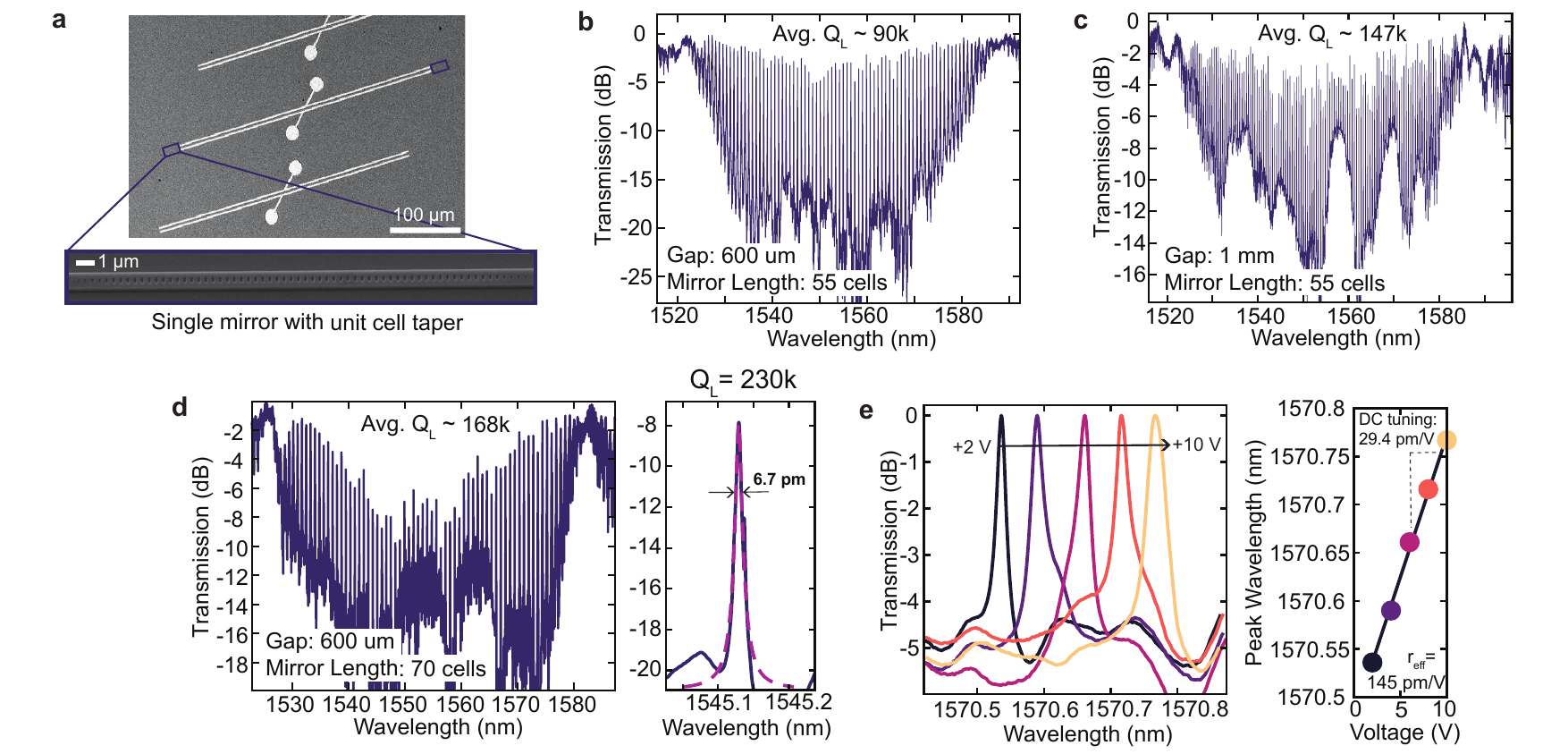}
    \caption{a) SEM of Fabry–Pérot cavities with deposited contacts; inlay describes individual FP mirror on either side of cavity. b) Example low-Q FP spectrum with a cavity gap of 600 microns and 55-mirror cell mirrors, average loaded Q of 90k, maximum loaded Q of 100k, and finesse of 50, demonstrating a maximum of 3dB power penalty. c) Example high-Q FP spectrum with a gap of 1 mm and 55 mirror cells, showing an average loaded Q of 147k, finesse of 53, and a maximum loaded Q exceeding 200k with a power penalty near the center of the bandgap of about 4.5 dB. d) FP spectrum from cavity with 600 micron length and 70 mirror cells with average loaded Q of 168k and finesse of 67; inlay describes maximum 230k Q peak chosen from near the bandgap center. e) FP tuning of resonant peak with DC tuning slope of 29.4 pm/V. 
}
    \label{fig:FIG3}
\end{figure}

On-chip Fabry--P\'{e}rot cavities can be fabricated on 1-D waveguides by etching two PhC mirrors a set distance apart on a straight waveguide, creating a resonant cavity within the bandgap of the surrounding mirrors (Figure \ref{fig:FIG3}a). Fabry--P\'{e}rot architectures have been extensively studied in Lithium Niobate~\cite{ref62,ref63}, as they enable a footprint reduction compared to ring and racetrack resonators, better control of modulation bandwidth design (through tuning $Q$-factor with mirror reflectivity), and can be operated in both transmission and reflection. 
With our photonic crystal mirrors, we fabricate Fabry--P\'{e}rot cavities with a range of $Q$-factors by controlling the reflectivity of our photonic crystal mirrors and the length of the cavity\cite{ref43}. We achieve a range of average $Q$-factors from 90k (Figure \ref{fig:FIG3}b) to 168k (Figure \ref{fig:FIG3}d), as calculated from the center 20 nm of each bandgap where reflectivity is highest. The systematic scaling of average Q with mirror design parameters confirms that cavity confinement is governed by mirror reflectivity rather than by scattering loss. If per-period scattering were dominant, Q would saturate independently of mirror count and cavity length; Figure \ref{fig:FIG3}b and Figure \ref{fig:FIG3}c demonstrate a clear increase in Q-factor across the cavity arising from an increase in cavity length, demonstrating that we are not dominated by intra-cavity loss. Additionally, we see the expected trend in cavity finesse, which reflects the tradeoff between Q-factor and cavity size \cite{ref43} between Figures \ref{fig:FIG3}b and \ref{fig:FIG3}c. These cavities have similar finesse values of 50 and 53, respectively, demonstrating the independence of finesse from cavity length. We see this trend continued in Figures \ref{fig:FIG3}b and \ref{fig:FIG3}d, where we see an increase in finesse (50 and 67 respectively) corresponding to an increase in mirror reflectivity with the same cavity length. We see resonances in the center of the bandgap reach a loaded $Q$ factor as high as 230k (Figure \ref{fig:FIG3}d) with a cavity length of 600 microns and 70 mirror cells, with an insertion loss of 8~dB and a finesse of 67. We perform DC tuning on FP resonances across the full $Q$ range under randomized bias between $+2$~V and $+10$~V (3$\times10^{5}$ V/m to 1.6$\times10^{6}$ V/m) (Figure \ref{fig:FIG3}e). The positive voltages here reflect bias applied only in the direction of poling to eliminate random depoling from our analysis (Supplementary Fig. 2a). We are able to extract the effective Pockels coefficient with the same resonant-modulator relation used for our racetracks, yielding $r_{\text{eff}} = 145$~pm/V with a DC tuning slope of 29.4 pm/V (interaction length of 600 microns) in agreement with the racetrack extracted values. 
\subsection{AC modulation of PhC-based devices}
\begin{figure}[htp]
    \centering
     \includegraphics[width=\textwidth]{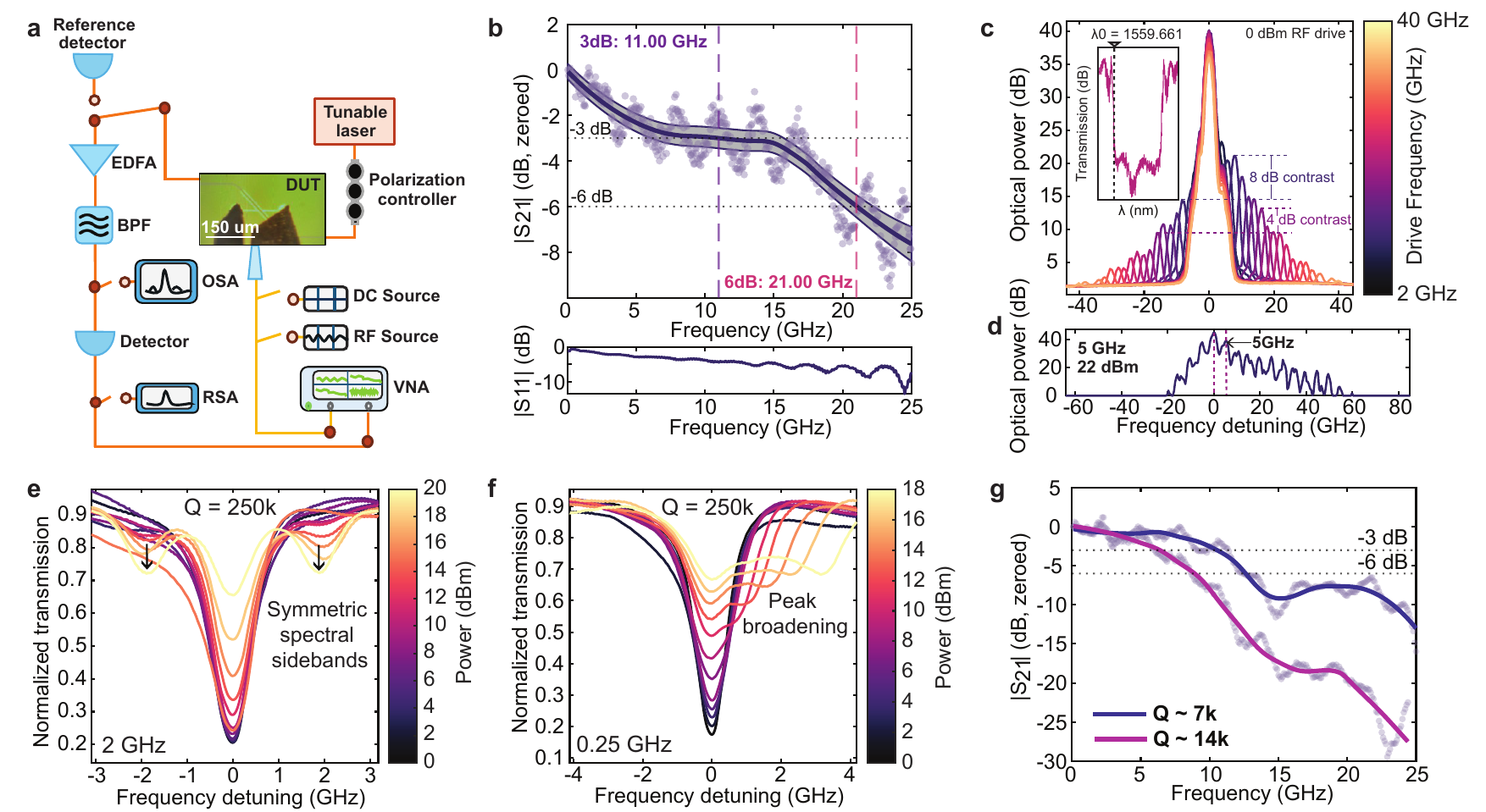}
    \caption{a) Modulation setup used for microwave and DC characterization.  b) Recorded scattering parameter S\textsubscript{21} from 100 MHz to 25 GHz, demonstrating a 3dB bandwidth of about 11 GHz and a 6dB bandwidth of 21 GHz (smoothed with a LOESS fit to account for oscillations arising from contact response). The lower inlay represents the contact response, S\textsubscript{11}, indicating a significant amount of reflection from contacts. c) Optical spectrum analyzer spectra at varying drive frequencies at 0 dBm of RF power demonstrating asymmetric sideband intensities on a photonic crystal bandedge. d) Intense generation of asymmetric sidebands at 22 dBm of driving RF power at 5 GHz e) Detuning of racetrack peak from modulation frequency less than the cavity line width ($Q_L$= 250k, modulation signal = 0.25 GHz) showing the formation of symmetric sidebands f) Visible tuning of sideband formation from modulation frequencies larger than cavity line width (Q = 250k, modulation signal = 2 GHz) g) Comparison of two different FP resonant peaks with different quality factors, demonstrating difference in 3dB bandwidth depending on cavity linewidth }
    \label{fig:FIG4}
\end{figure}
Having established the optical quality and DC electro-optic response of our BTO photonic crystals, we turn to characterizing their AC modulation performance. A schematic of our characterization system is shown in Figure \ref{fig:FIG4}a and described in depth in the methods section. A tunable laser is used with a polarization controller to couple to the device under test. For bandwidth measurements, we use a DC source and vector network analyzer to characterize the modulation of our devices. To first characterize the response of the underlying film, we use a non-resonant device in the form of a photonic crystal. The band edge has an asymmetric transmission response, yielding a strong modulation in transmission when modulated by a vector network analyzer (VNA). This bandwidth is determined by the electro-optic response of the BTO material itself, rather than the photon lifetime of a resonant cavity. This configuration is  potentially useful for the development of slow light modulators \cite{PhCSlowLightModulator}. 

Our device under test uses an electrode thickness of 200 nm and contact spacing of 6~\textmu m, yielding a simulated $\Gamma_{\text{eo}} = 0.51$ (see Supplementary Fig. 3). We modulate the band edge and extract a 3-dB electro-optic bandwidth of approximately 11~GHz, and a 6-dB electro-optic bandwidth of about 21~GHz (Figure \ref{fig:FIG4}b) while driving at a power of -2 dBm at a wavelength of 1559.661 nm. We do not observe higher order sidebands generated using this power as measured on our optical spectrum analyzer. Comparing our measured $S_{21}$ roll-off with the frequency-dependent $r_{42}$ reported in recent broadband dielectric characterization~\cite{ref56} shows a similar roll-off of $r_{42}$ that begins near 10~GHz. Further improvements in contact design can reduce the observed ringing from impedance mismatch and potentially improve the bandwidth further (see Supplementary Fig. 4). 

The dispersion of the PhC band edge introduces an additional nonlinearity that is absent for modulation on a symmetric Lorentzian resonance. Sidebands above and below the injected center frequency are unequally filtered through the photonic crystal, yielding a quasi-single sideband modulation~\cite{ref55,pohl2020100}. Using an RF signal generator rather than a vector network analyzer, we characterize this in Figure \ref{fig:FIG4}c, which shows the measured sideband generation with a 0 dBm RF power as a function of frequency. We observe a sideband asymmetry of 6-8 dB. This contrasts with the symmetric, lifetime-limited sideband cascade we observe in resonant cavities. When driven at high powers (>20 dBm), sharp BTO PhC band edges generate an asymmetric frequency-comb-like spectra (Figure \ref{fig:FIG4}d) with asymmetric higher-order sideband formation. We observe 15 comb lines at this power, which can be improved with further improvements to our fabrication and electrode design. 

Finally, we characterize the modulation performance of resonant cavities. Sweeping the laser wavelength across a racetrack resonator resonance of linewidth 0.0062 nm ($Q_L = 250$k) while applying a sinusoidal RF tone, we observe two distinct regimes (Figure \ref{fig:FIG4}e, Figure \ref{fig:FIG4}f). When the RF frequency (chosen at 2~GHz) is larger than the resonance linewidth (which corresponds to 773.3 MHz) (Figure \ref{fig:FIG4}e), the resonance cannot respond quickly enough, and modulation sidebands form at integer multiples of the RF frequency around the carrier.  When the RF frequency is smaller than the linewidth (chosen at 0.25 GHz, \ref{fig:FIG4}f), the resonance follows the modulation in real time and the transmission peak appears as a broadened doublet. Due to BTO's thermo-optic coefficient, at powers over 6~dBm we can see a slight shift in spectral location of our peak during test, as demonstrated in  (Figure \ref{fig:FIG4}f)~\cite{ref64}. 

Additionally, we explore the bandwidth limited response of Fabry–Pérot cavities, as shown in Figure \ref{fig:FIG4} g, where the dark purple line represents the RF response of a resonance with a $Q$ of ~7,000, and the magenta line represents the RF response of a resonance with a $Q$ of 14,000. We see that the 3~dB bandwidth of the two resonances ($\sim$11~GHz, $\sim$6~GHz) scales with the $Q$ factor. Although these resonances still show ringing and a significant contact response, the linear scaling suggests that their response is still dominated by cavity linewidth. Because contact-limited bandwidth would be set by electrode geometry alone and therefore independent of cavity Q, this scaling instead indicates that the resonant response here is governed by cavity photon lifetime and material response rather than by contact behavior.

\subsection{Discussion and outlook}
This work has demonstrated a viable route to high-performance sub-wavelength nanostructures in monolithic BTO on insulator. The high modulation bandwidth of the material, combined with the dielectric confinement allowed by photonic crystals, open up new possibilities for dispersion engineering and deeply subwavelength control of light on-chip. The enhanced electro-optic coefficient and monolithic nature of our BTOI devices promises simultaneously compact and energy-efficient optical modulators, amenable for future computing paradigms that require ultra low switching energy and high integration density~\cite{shastri2021photonics,jha2022nanophotonic}. The demonstrated feature size of these BTO nanostructures opens up the possibility to create visible frequency modulators with high fidelity, opening potential for use cases in quantum sensing \cite{VisibleLightQuantum}, where emission from qubits is often in the visible spectrum. The dispersion engineering inherent in this approach may also be useful in creating nonlinear frequency combs\cite{yu2019photonic}. As we move to further investigate BTO as an integrated photonic platform, we can explore how to leverage BTO's ferroelectric domain behavior through the lens of nanostructures. For example, BTO's soft ferroelectric domain behavior poses opportunities around device re-programmability \cite{BTOhighspeedarray} and non-volatile memory applications \cite{BTOnonvolatilephaseshifter}. Because non-volatile domain writing produces only a small phase shift and energetically scales with the volume of material poled \cite{BTOnonvolatilephaseshifter}, leveraging BTO nanostructures reduces the energy per write operation, as well as enhances the light-matter interaction for the readout of a non-volatile phase shift. The ferroelectric domain structure and nonlinear polarization are strongly dependent on growth conditions and strain engineering; further improvements to BTO growth by controlling stoichiometry \cite{cavanagh2025effect} or strain \cite{deng2026self} may further improve the electro-optic strength or bandwidth, providing a route to ultra-efficient integrated photonics in a CMOS-compatible platform. 

\subsection{Methods}
\subsubsection{Device design and fabrication}
Single mode waveguides and photonic crystals are designed using commercially available software (ANSYS Lumerical and COMSOL Multiphysics). Guided modes are simulated using Lumerical's eigenmode solver to confirm single mode operation for TE modes and to estimate the mode refractive index. The photonic crystal bandstructure is computed using COMSOL Multiphysics, and the transmission response of these photonic crystals are computationally validated using Lumerical FDTD simulations. All simulations are convergence tested by modifying the mesh density and domain size. 

All devices were fabricated using standard nanofabrication techniques. First, a negative-tone resist (ma-N 2405) was spun onto the sample at 500 nm thickness and baked at 90 °C before being patterned using a 50 kV electron beam lithography system (Raith Voyager). Following development in commercial developer (AZ917), the devices were etched using inductively coupled plasma reactive ion etching (ULVAC NLD-5700). Our etch employs both physical (Ar) and chemical (CHF$_3$) etch gases, and the etching conditions are optimized to minimize mask erosion while maintaining etch anisotropy. Our optimized process yields an etch rate of approximately 12 nm/min. Following this, we use a multi-step cleaning procedure to remove etch redeposition and the mask, including oxygen plasma descumming, resist removal in solvents (NMP 1165, acetone, and IPA), and a KOH bath (see Supplementary Fig. 1). These steps are developed to minimize chemical etch roughness that can be induced from chemical redeposition removal. After redeposition removal, our devices are annealed in an oxygen environment at a moderate temperature (500-600 °C) for several hours to improve the quality factor. Gold contacts are deposited (typically 200-300 nm) using a 4 nm chromium adhesion layer with photolithography (Heidelberg Maskless aligner) and electron beam evaporation (AJA). The electrode gap is 6 microns. Finally, the chip facets are cleaved for optical measurements. 

To pole our samples, we use a commercial high voltage supply (Radiant Technologies) and electrical probes (Signatone). 200 V is applied to the sample for 10 minutes before ramping down to 0V. The state of poling is evaluated using confocal scanning second harmonic generation microscopy using a custom-built system based on a Thorlabs Bergamo microscope. A FemtoFiber ultra 920 ultrafast fiber laser system (Toptica) is used, which features bandwidth limited 120 fs optical pulses with a center wavelength of 920 nm and repetition rate of 80 MHz. Optical pulses are delivered to the Thorlabs Bergamo confocal microscope using a dispersion-compensated hollow core fiber, and the polarization is set using a linear polarizer and half wave plate. Polarization-resolved second harmonic generation images are then acquired using a long working distance 50X objective (NA 0.7, Thorlabs) and the second harmonic signal is collected using a photomultiplier tube following a dichroic filter (Thorlabs). Intensity is used as a proxy for state of poling.

SEM images are collected using a JEOL JIB-4700F scanning electron microscope, typically using an accelerating voltage of 3 kV and a probe current setting of 8, which corresponds to an electron beam current between 0.1 nA and 0.5 nA.  

\subsubsection{Propagation Loss}
To calculate our propagation loss values, we use \cite{ref19}: 
\begin{equation}
    \alpha   (dB/cm)= \frac{2\cdot\pi \cdot n_g}{10^{-11} \cdot Q_i \cdot \lambda}\cdot10\text{log}_{10} e
\end{equation}
where $n_g$ is the group index and $\lambda$ is the average wavelength (nm) over which the $Q_i$ values are drawn from. The shaded region in (Figure \ref{fig:FIG1}d) represents the confidence interval of the linear fit. 

\subsubsection{Optical Characterization Setup}
To characterize our devices, we use a tunable semiconductor laser (Santec TSL-570) and polarization controller (Thorlabs) to excite the fundamental TE mode of our single mode waveguides. Light is coupled onto the chip via fiber-to-fiber system using lensed optical fibers (Oz optics). Our fiber-to-fiber insertion loss typically is about 12~dB. We measure transmission spectra using a Santec MPM-220 optical power meter. Quality factors are determined by fitting the resulting resonance spectra to a Lorentzian function. For DC modulation of our device under test (DUT), we use a PicoProbe Model 40A microwave probe with 150 $\mu$m pitch and a DC power source (Keysight E36313A) to apply DC voltages up to 50~V. Effective electro-optic strength measurements are performed by measuring transmission spectra as a function of applied voltage. BTO is a soft ferroelectric, meaning that sufficiently high voltages can begin to depole our devices. We use a sufficiently low voltage range to minimize this, and additionally characterized the response of our devices when we randomly set the voltage, rather than sweeping from low to high DC voltages.

Our probe is connected to a 50~GHz bias tee that allows us to apply DC and microwave power simultaneously. To measure the microwave bandwidth using photonic crystals and the resonant bandwidth limited response of our optical cavities, we use a Rohde and Schwarz vector network analyzer (VNA), with powers ranging from -2 dBm to 0 dBm. First, a transmission spectrum is recorded and the laser wavelength is set near the band edge of a photonic crystal. The modulated signal at the output of the photonic crystal is collected using a lensed fiber, amplified using an erbium-doped fiber amplifier (EDFA, FiberProme Model: EDFA-C26G-S) and a 1-nm-width bandpass filter (Newport TBF-1550-1.0) to remove unwanted amplified spontaneous emission before being measured by a high speed photodetector (Newport Model 1014) and connected to the receiver port of the VNA to measure the S21 response. 

To measure sideband asymmetry and frequency comb generation, we instead use a Rohde and Schwarz radio frequency generator (SMA100B) to apply an RF tone and measure the resulting optical spectrum using a Yokogawa optical spectrum analyzer (AQ6370E). Finally, we measure the optical response of optical cavities in the sideband resolved and sideband unresolved regimes using the Rohde and Schwarz radio frequency generator (SMA100B), but measure the resulting spectrum by sweeping the laser wavelength and recording the transmitted power using a photodetector as a function of RF power.

\newpage
%%=========================================================
\section{Acknowledgments}
%%=========================================================
We would like to thank Dr. Serkan Butun at Northwestern, as well as Dr. Ralu Divan and Dr. Dave Czaplewski at Argonne National Lab for their discussions on improving electron beam lithography performance. We additionally would like to thank Diego De La Vega and Dr. Shaoning Lu for their work on the ULVAC etcher. Financial support for this work was provided by the National Science Foundation Graduate Research Fellowship under Grant No.\ DGE-2234667. The authors would additionally like to thank Komal Prasad and Abhiram Devata for discussions around Fabry–Pérot design and Lumerical simulation, respectively. This work made use of the NUFAB facility (RRID:SCR\_017779) of Northwestern University’s NUANCE Center, which has received support from the IIN and Northwestern’s MRSEC program (NSF DMR-2308691).

\end{spacing}

%%===================================================================
\section{Author information}
%%===================================================================
\subsubsection{Authors and Affiliations}
Department of Materials Science and Engineering, McCormick School of Engineering, Northwestern University, 2220 Campus Drive, Evanston, IL, 60208:
\textbf{Sarah Berman, Sina Dereshgi, David Barton}

\subsubsection{Contributions}
D.B. and S.B. designed the project. S.B. designed and simulated the devices, developed the fabrication protocols and fabricated the devices, characterized the devices, and analyzed the data. S.D. and S.B. setup and calibrated the AC and DC modulation setup. S.D. poled the devices, S.B. and S.D. performed SHG.  D.B. supervised the work. All authors discussed the results and commented on the manuscript.
\printbibliography

@article{ref14,
  title={A {BaTiO$_3$}-Based Electro-Optic Pockels Modulator Monolithically Integrated on an Advanced Silicon Photonics Platform},
  author={Eltes, F. and Mai, C. and Caimi, D. and Kroh, M. and Popoff, Y. and Winzer, G. and Petousi, D. and Lischke, S. and Ortmann, J. E. and Czornomaz, L. and Zimmermann, L. and Fompeyrine, J. and Abel, S.},
  journal={Journal of Lightwave Technology},
  volume={37},
  number={5},
  pages={1456--1462},
  year={2019}
}

@article{ref17,
  title={Low loss monolithic barium titanate on insulator integrated photonics with intrinsic quality factor $>$1 million},
  author={Kim, G. I. and Yim, J. and Bahl, G.},
  journal={arXiv preprint arXiv:2507.17150},
  doi={10.48550/arXiv.2507.17150},
  year={2025}
}

@article{ref19,
  title={Monolithic Barium Titanate Modulators on Silicon-on-Insulator Substrates},
  author={Dong, Z. and Raju, A. and Posadas, A. B. and Reynaud, M. and Demkov, A. A. and Wasserman, D. M.},
  journal={ACS Photonics},
  volume={10},
  number={12},
  pages={4367--4376},
  year={2023}
}

@article{ref30,
  title={Ultra-Low-Power Tuning in Hybrid Barium Titanate--Silicon Nitride Electro-optic Devices on Silicon},
  author={Ortmann, J. E. and Eltes, F. and Caimi, D. and Meier, N. and Demkov, A. A. and Czornomaz, L. and Fompeyrine, J. and Abel, S.},
  journal={ACS Photonics},
  volume={6},
  number={11},
  pages={2677--2684},
  year={2019}
}

@article{ref39,
  title={High-{Q} Monolithic Ring Resonators in Low-Loss Barium Titanate on Silicon},
  author={Raju, A. and Hungund, D. and Krueger, D. and Dong, Z. and Sakotic, Z. and Posadas, A. B. and Demkov, A. A. and Wasserman, D.},
  journal={Laser \& Photonics Reviews},
  volume={19},
  number={16},
  pages={2402086},
  year={2025}
}

@book{ref41,
  title={Photonic Crystals: Molding the Flow of Light},
  author={Joannopoulos, J. D. and Johnson, S. G. and Winn, J. N. and Meade, R. D.},
  edition={2nd},
  publisher={Princeton University Press},
  address={Princeton},
  year={2008}
}

@article{ref43,
  title={High-{Q} and compact {Fabry--P\'{e}rot} microresonators on thin-film lithium niobate},
  author={Yang, L. and Li, C. and Xie, J. and Tang, H. X.},
  journal={Nanophotonics},
  volume={14},
  number={26},
  pages={4675--4681},
  year={2025}
}

@article{ref45,
  title={Lithium niobate photonic-crystal electro-optic modulator},
  author={Li, M. and Ling, J. and He, Y. and Javid, U. A. and Xue, S. and Lin, Q.},
  journal={Nature Communications},
  volume={11},
  pages={4123},
  year={2020}
}

@article{ref51,
  title={Two-Dimensional Ferroelectric Photonic Crystal Waveguides: Simulation, Fabrication, and Optical Characterization},
  author={Lin, P. T. and Yi, F. and Ho, S.-T. and Wessels, B. W.},
  journal={Journal of Lightwave Technology},
  volume={27},
  number={19},
  pages={4330--4337},
  year={2009}
}

@article{ref52,
  title={Silicon Photonics Circuit Design: Methods, Tools and Challenges},
  author={Bogaerts, W. and Chrostowski, L.},
  journal={Laser \& Photonics Reviews},
  volume={12},
  number={4},
  pages={1700237},
  year={2018}
}

@article{ref55,
  title={Mono-drive single-sideband modulation via optical delay lines on thin-film lithium niobate},
  author={Chen, Y. and Feng, H. and Wang, Z. and Zhang, K. and Xie, X. and Zeng, Y. and Ren, Y. and Wang, C.},
  journal={Optica},
  volume={12},
  number={5},
  pages={666--673},
  year={2025}
}

@article{ref56,
  title={Barium titanate and lithium niobate permittivity and {Pockels} coefficients from megahertz to sub-terahertz frequencies},
  author={Chelladurai, D. and Kohli, M. and Winiger, J. and Moor, D. and Messner, A. and Fedoryshyn, Y. and Eleraky, M. and Liu, Y. and Wang, H. and Leuthold, J.},
  journal={Nature Materials},
  volume={24},
  pages={868--875},
  year={2025}
}

@inproceedings{ref57,
  title={Electro-Optic Barium Titanate Modulators on Silicon Photonics Platform},
  author={Posadas, A. B. and Stenger, V. E. and DeFouw, J. D. and Warner, J. H. and Demkov, A. A.},
  booktitle={2023 IEEE Silicon Photonics Conference (SiPhotonics)},
  pages={1--2},
  year={2023},
  doi={10.1109/SiPhotonics55903.2023.10141930}
}

@article{ref58,
  title={Propagation losses of silicon nitride waveguides in the near-infrared range},
  author={Melchiorri, M. and Daldosso, N. and Sbrana, F. and Pavesi, L. and Pucker, G. and Kompocholis, C. and Bellutti, P. and Lui, A.},
  journal={Applied Physics Letters},
  volume={86},
  pages={121111},
  year={2005}
}

@article{ref59,
  title={Integrated photonics on thin-film lithium niobate},
  author={Zhu, D. and Shao, L. and Yu, M. and Cheng, R. and Desiatov, B. and Xin, C.J. and Hu, Y. and Holzgrafe, J. and Ghosh, S. and Shams-Ansari, A. and Puma, E. and Sinclair, N. and Reimer, C. and Zhang, M. and Lon{\v{c}}ar, M.},
  journal={Advances in Optics and Photonics},
  volume={13},
  number={2},
  pages={242--352},
  year={2021}
}

@article{ref61,
  title={High extinction ratio bandgap of photonic crystals in {LNOI} wafer},
  author={Zhang, S.-M. and Cai, L.-T. and Jiang, Y.-P. and Jiao, Y.},
  journal={Optical Materials},
  volume={64},
  pages={203--207},
  year={2017}
}

@article{ref62,
  title={Ultra-compact lithium niobate photonic chip for high-capacity and energy-efficient wavelength-division-multiplexing transmitters},
  author={Liu, H. and Pan, B. and Li, H. and Yu, Z. and Liu, L. and Shi, Y. and Dai, D.},
  journal={Light: Advanced Manufacturing},
  volume={4},
  pages={133--142},
  year={2023}
}

@article{ref63,
  title={Tunable Ultranarrowband Grating Filters in Thin-Film Lithium Niobate},
  author={Prencipe, A. and Baghban, M. A. and Gallo, K.},
  journal={ACS Photonics},
  volume={8},
  number={10},
  pages={2923--2930},
  year={2021}
}

@article{ref64,
  title={Thermo-optic coefficient of single-crystal thin-film barium titanate on insulator},
  author={Lin, H.-L. and Yin, Y. and Roos, L. and Dogheche, E. and Danner, A. J.},
  journal={Optical Materials Express},
  volume={15},
  number={12},
  pages={2975--2983},
  year={2025}
}

@article{LithiumNiobatePhotonics,
  title={Lithium niobate photonics: Unlocking the electromagnetic spectrum},
  author={Boes, A. and Chang, L. and Langrock, C. and Yu, M. and Zhang, M. and Lin, Q. and Lon{\v{c}}ar, M. and Fejer, M. and Bowers, J. and Mitchell, A.},
  journal={Science},
  volume={379},
  number={6627},
  pages={eabj4396},
  year={2023}
}

@article{CapacitanceCitation,
  title={Review and perspective on ultrafast wavelength-size electro-optic modulators},
  author={Liu, K. and Ye, C. R. and Khan, S. and Sorger, V. J.},
  journal={Laser \& Photonics Reviews},
  volume={9},
  number={2},
  pages={172--194},
  year={2015}
}

@article{MicrowaveIntro,
  title={Integrated microwave photonics},
  author={Marpaung, D. and Yao, J. and Capmany, J.},
  journal={Nature Photonics},
  volume={13},
  pages={80--90},
  year={2019}
}

@article{AdvancedComputeIntro,
  title={Experimental demonstration of reservoir computing on a silicon photonics chip},
  author={Vandoorne, K. and Mechet, P. and Van Vaerenbergh, T. and Fiers, M. and Morthier, G. and Verstraeten, D. and Schrauwen, B. and Dambre, J. and Bienstman, P.},
  journal={Nature Communications},
  volume={5},
  pages={3541},
  year={2014}
}

@article{QuantumIntro,
  title={Programmable integrated quantum photonics},
  author={Aharonovich, I. and Crozier, K. B. and Neshev, D.},
  journal={Nature Photonics},
  volume={20},
  pages={254--265},
  year={2026}
}

@article{OpticalCommIntro,
  title={Integrated photonics enabling ultra-wideband fibre--wireless communication},
  author={Zhang, Y. and Shu, H. and Guo, Y. and Zhou, P. and Wang, L. and others},
  journal={Nature},
  year={2026}
}

@article{LNNanoStructures1,
  title={Second harmonic generation in nano-structured thin-film lithium niobate waveguides},
  author={Wang, C. and Xiong, X. and Andrade, N. and Venkataraman, V. and Ren, X.-F. and Guo, G.-C. and Lon{\v{c}}ar, M.},
  journal={Optics Express},
  volume={25},
  number={6},
  pages={6963--6973},
  year={2017}
}

@article{LNNanoStructures2,
  title={Ultralow-Threshold Lithium Niobate Photonic Crystal Nanocavity Laser},
  author={Liu, X. and Chen, C. and Ge, R. and Wu, J. and Chen, X. and Chen, Y.},
  journal={Nano Letters},
  volume={25},
  number={16},
  pages={6454--6460},
  year={2025}
}

@article{SiWaveguide,
  title={Single Virus Detection on Silicon Photonic Crystal Random Cavities},
  author={Watanabe, K. and Wu, H.-Y. and Xavier, J. and Joshi, L. T. and Vollmer, F.},
  journal={Small},
  volume={18},
  number={15},
  pages={2107597},
  year={2022}
}

@article{GaAsWaveguide,
  title={Hybrid {Si-GaAs} photonic crystal cavity for lasing and bistability},
  author={Rahaman, M. H. and Lee, C.-M. and Buyukkaya, M. A. and Zhao, Y. and Waks, E.},
  journal={Optics Express},
  volume={31},
  number={23},
  pages={37574--37582},
  year={2023}
}

@article{SiNWaveguide,
  title={Photonic crystal waveguides on silicon rich nitride platform},
  author={Debnath, K. and Dominguez Bucio, T. and Al-Attili, A. and Khokhar, A. Z. and Saito, S. and Gardes, F. Y.},
  journal={Optics Express},
  volume={25},
  number={4},
  pages={3214--3221},
  year={2017}
}

@article{PolarizationMultiplexer,
  title={Compact Polarization-Multiplexed Optical Transmitters through the Monolithic Integration of Metasurface and Photonic Crystal Surface-Emitting Lasers},
  author={Miao, W.-C. and Hsiao, F.-H. and Chang, Y.-H. and Jhang, Y.-H. and Cheng, C.-H. and Yu, H.-C. and Lin, G.-R. and Lin, C.-L. and Chow, C.-W. and Hong, Y.-H. and Kuo, H.-C.},
  journal={ACS Photonics},
  volume={12},
  number={8},
  pages={4432--4439},
  year={2025}
}

@article{TopologicalEndStates,
  title={Broadband localization of light at the termination of a topological photonic waveguide},
  author={Muis, D. and Li, Y. and Barczyk, R. and Arora, S. and Kuipers, L. and Shvets, G. and Verhagen, E.},
  journal={Science Advances},
  volume={11},
  number={16},
  pages={eadr9569},
  year={2025}
}

@article{FPFreqComb,
  title={Electro-optic frequency comb generation in lithium niobate photonic crystal {Fabry--P\'{e}rot} micro-resonator},
  author={Hwang, H. and Go, S. and Kim, G. and others},
  journal={npj Nanophotonics},
  volume={3},
  pages={15},
  year={2026}
}

@article{BTOhighspeedarray,
  title={High-speed non-volatile barium titanate field-programmable photonic gate array},
  author={Catal{\'a}-Lahoz, C. and Rausell-Campo, J. R. and P{\'e}rez-L{\'o}pez, D. and G{\"u}niat, L. and Convertino, C. and Eltes, F. and Fompeyrine, J. and Mesaritakis, C. and Bogris, A. and Wilmart, Q. and Faugier-Tovar, J. and Charbonnier, B. and Capmany, J.},
  journal={Nature Photonics},
  year={2026}
}

@article{BTOnonvolatilephaseshifter,
  title={A ferroelectric multilevel non-volatile photonic phase shifter},
  author={Geler-Kremer, J. and Eltes, F. and Stark, P. and Stark, D. and Caimi, D. and Siegwart, H. and Offrein, B. J. and Fompeyrine, J. and Abel, S.},
  journal={Nature Photonics},
  volume={16},
  pages={491--497},
  year={2022}
}

@article{VisibleLightQuantum,
  title={Quantum Photonics on a Chip},
  author={Katiyi, A. and Karabchevsky, A.},
  journal={APL Quantum},
  volume={2},
  number={2},
  pages={020901},
  year={2025}
}

@article{QuantumEmitter,
  title={Photonic-Crystal-Enhanced Single-Photon Sources Based on Semiconductor Quantum Dots},
  author={Wang, Y. and Mu, Y. and Li, Y. and Gao, F. and Liu, F.},
  journal={Chip},
  year={2026}
}

@article{SlowlightPhCs,
  title={Slow light in photonic crystals},
  author={Baba, T.},
  journal={Nature Photonics},
  volume={2},
  number={8},
  pages={465--473},
  year={2008}
}

@article{WhySlowLight,
  title={Why do we need slow light?},
  author={Krauss, T. F.},
  journal={Nature Photonics},
  volume={2},
  number={8},
  pages={448--450},
  year={2008}
}

@article{PhCSlowLightModulator,
  title={Enhancement of the pockels effect in photonic crystal modulators through slow light},
  author={Girouard, P. and Liu, Z. and Chen, P. and Jeong, Y. K. and Tu, Y. and Ho, S.-T. and Wessels, B. W.},
  journal={Optics Letters},
  volume={41},
  number={23},
  pages={5531--5534},
  year={2016}
}

@article{yu2022integrated,
  title={Integrated femtosecond pulse generator on thin-film lithium niobate},
  author={Yu, M. and Barton III, D. and Cheng, R. and Reimer, C. and Kharel, P. and He, L. and Shao, L. and Zhu, D. and Hu, Y. and Grant, H. R. and others},
  journal={Nature},
  volume={612},
  number={7939},
  pages={252--258},
  year={2022},
  publisher={Nature Publishing Group UK London}
}

@article{du2020silicon,
  title={Silicon nitride chirped spiral Bragg grating with large group delay},
  author={Du, Z. and Xiang, C. and Fu, T. and Chen, M. and Yang, S. and Bowers, J. E. and Chen, H.},
  journal={APL Photonics},
  volume={5},
  number={10},
  year={2020},
  publisher={AIP Publishing}
}

@article{ohta2011strong,
  title={Strong coupling between a photonic crystal nanobeam cavity and a single quantum dot},
  author={Ohta, R. and Ota, Y. and Nomura, M. and Kumagai, N. and Ishida, S. and Iwamoto, S. and Arakawa, Y.},
  journal={Applied Physics Letters},
  volume={98},
  number={17},
  year={2011},
  publisher={AIP Publishing}
}

@article{ding2025purcell,
  title={Purcell-enhanced emissions from diamond color centers in slow light photonic crystal waveguides},
  author={Ding, S. W. and Jin, C. and Kuruma, K. and Guo, X. and Haas, M. and Korzh, B. and Beyer, A. and Shaw, M. D. and Sinclair, N. and Awschalom, D. D. and others},
  journal={Nano Letters},
  volume={25},
  number={32},
  pages={12125--12131},
  year={2025},
  publisher={ACS Publications}
}

@article{witmer2020silicon,
  title={A silicon-organic hybrid platform for quantum microwave-to-optical transduction},
  author={Witmer, J. D. and McKenna, T. P. and Arrangoiz-Arriola, P. and Van Laer, R. and Alex Wollack, E. and Lin, F. and Jen, A. K.Y. and Luo, J. and Safavi-Naeini, A. H.},
  journal={Quantum Science \& Technology},
  volume={5},
  number={3},
  pages={034004},
  year={2020},
  publisher={IOP Publishing}
}

@article{shastri2021photonics,
  title={Photonics for artificial intelligence and neuromorphic computing},
  author={Shastri, B. J. and Tait, A. N. and Ferreira de Lima, T. and Pernice, W. H.P. and Bhaskaran, H. and Wright, C. D. and Prucnal, P. R.},
  journal={Nature Photonics},
  volume={15},
  number={2},
  pages={102--114},
  year={2021},
  publisher={Nature Publishing Group UK London}
}

@article{jha2022nanophotonic,
  title={Nanophotonic cavity based synapse for scalable photonic neural networks},
  author={Jha, A. and Huang, C. and Delima, T. F. and Peng, H.-T. and Shastri, B. and Prucnal, P. R.},
  journal={IEEE journal of selected topics in quantum electronics},
  volume={28},
  number={6: High Density Integr. Multipurpose Photon. Circ.},
  pages={1--8},
  year={2022},
  publisher={IEEE}
}

@article{yu2019photonic,
  title={Photonic-crystal-reflector nanoresonators for Kerr-frequency combs},
  author={Yu, S.-P. and Jung, H. and Briles, T. C. and Srinivasan, K. and Papp, S. B.},
  journal={ACS Photonics},
  volume={6},
  number={8},
  pages={2083--2089},
  year={2019},
  publisher={ACS Publications}
}

@article{cavanagh2025effect,
  title={Effect of Stoichiometry on the Structure and Polarization of BaTiO3},
  author={Cavanagh, A. E. and Little, L. B. and Zhang, Y. and Barton, D. R. and Taylor, N. K. and Lin, H.-Y. G. and Butler, A. and N'Diaye, A. T. and Gardener, J. A. and Brooks, C. M. and others},
  journal={Advanced Optical Materials},
  volume={13},
  number={22},
  pages={2501037},
  year={2025},
  publisher={Wiley Online Library}
}

@article{deng2026self,
  title={Self-buffered epitaxy of barium titanate on oxide insulators enables high-performance electro-optic modulators},
  author={Deng, C. and He, Y. and Yang, W. and Yu, H. and Hong, Z. and Liu, H. and Han, H. and Li, W. and Ma, Y. and Zhang, Z. and others},
  journal={Light: Science \& Applications},
  volume={15},
  number={1},
  pages={21},
  year={2026},
  publisher={Nature Publishing Group UK London}
}

@article{prountzou2026electro,
  title={Electro-optic Modulation in Polycrystalline Barium Titanate Metasurfaces Enhanced by Poling},
  author={Prountzou, E. and Weigand, H. C. and Falcone, V. and Talts, U.-L. and Trassin, M. and Grange, R.},
  journal={ACS photonics},
  volume={13},
  number={10},
  pages={2928--2936},
  year={2026},
  publisher={ACS Publications}
}

@article{PhotonicsForComputing,
  title={Industry insight: photonics to scale {AI} data centers},
  author={Torrijos-Mor{\'a}n, L. and P{\'e}rez-L{\'o}pez, D.},
  journal={npj Nanophotonics},
  volume={3},
  number={1},
  pages={8},
  year={2026},
  doi={10.1038/s44310-025-00105-1}
}

@article{pohl2020100,
  title={100-GBd waveguide Bragg grating modulator in thin-film lithium niobate},
  author={Pohl, David and Messner, Andreas and Kaufmann, Fabian and Escale, Marc Reig and Holzer, Jannis and Leuthold, Juerg and Grange, Rachel},
  journal={IEEE Photonics Technology Letters},
  volume={33},
  number={2},
  pages={85--88},
  year={2020},
  publisher={IEEE}
}

\end{document}